\documentclass[conference]{IEEEtran}
\IEEEoverridecommandlockouts
\usepackage{cite}
\usepackage{amsmath,amssymb,amsfonts}
\usepackage{caption}
\usepackage{algorithmic}
\usepackage{graphicx}
\usepackage{textcomp}
\usepackage{xcolor}
\def\BibTeX{{\rm B\kern-.05em{\sc i\kern-.025em b}\kern-.08em
    T\kern-.1667em\lower.7ex\hbox{E}\kern-.125emX}}
\begin{document}

\title{
Custom 8-bit floating point value format for reducing shared memory bank conflict in approximate nearest neighbor search
\thanks{(*) This research is part of the contents on which Hiroyuki Ootomo worked in the NVIDIA internship.}
}
\author{\IEEEauthorblockN{1\textsuperscript{st} Hiroyuki Ootomo}
\IEEEauthorblockA{\textit{Dept. of Computer Science} \\
\textit{Tokyo Institute of Technology}\\
Tokyo, Japan \\
ootomo.h@rio.gsic.titech.ac.jp}
\and
\IEEEauthorblockN{2\textsuperscript{nd} Akira Naruse}
\IEEEauthorblockA{
\textit{NVIDIA}\\
Tokyo, Japan \\
anaruse@nvidia.com}
}

\maketitle

\begin{abstract}
The $k$-nearest neighbor search is used in various applications such as machine learning, computer vision, database search, and information retrieval.
While the computational cost of the exact nearest neighbor search is enormous, an approximate nearest neighbor search (ANNS) has been attracting much attention.
IVFPQ is one of the ANNS methods.
Although we can leverage the high bandwidth and low latency of shared memory to compute the search phase of the IVFPQ on NVIDIA GPUs, the throughput can degrade due to shared memory bank conflict. 
To reduce the bank conflict and improve the search throughput, we propose a custom 8-bit floating point value format.
This format doesn't have a sign bit and can be converted from/to FP32 with a few instructions.
We use this format for IVFPQ on GPUs and achieved better performance without significant recall loss compared to FP32 and FP16.
\end{abstract}

\begin{IEEEkeywords}
Approximate Nearest Neighbor Search, IVFPQ, Floating point value
\end{IEEEkeywords}

\section{Introduction}
%
\subsection{NVIDIA GPU architecture and shared memory}
In NVIDIA GPUs, 32 threads in a warp works concurrently.
The shared memory is an on-chip memory that has higher bandwidth, lower latency, and smaller capacity compared to device memory.
This memory has 32 banks where each one has 4 bytes of data width and the threads access the shared memory through them.
When all threads access different banks, they are executed in parallel.
However, when some threads in a warp access different addresses but using the same bank, they are executed sequentially. 
This is called ``bank conflict" and should be avoided to get higher performance.

\subsection{Approximate nearest neighbor search}
The $k$-nearest neighbor search ($k$-NNS) seeks $k$ vectors $V=\{\mathbf{x}_{i_1}, \mathbf{x}_{i_2}, \cdots, \mathbf{x}_{i_k}\}$ for a given query vector $q$ from a dataset $D=\{\mathbf{x}_1, \mathbf{x}_2, \cdots, \mathbf{x}_N\}$ where
\begin{equation*}
    i_1, i_2, \cdots, i_k = \text{k-argmins}_i \left( |\mathbf{q} - \mathbf{x}_i |^2\right).
\end{equation*}
Each vector is $d$ dimensional.
The $k$-approximate nearest neighbor search computes $k$-NNS approximately, which means the result vectors are not always top-$k$ nearest.
We evaluate the accuracy of $k$-ANNS by calculating the ``recall" as follows:
\[
    \text{recall} = |V \cap V_{GT}| / |V_{GT}|,
\]
where $V_\text{GT}$ is a set of ground truth vectors.

\subsection{IVFPQ algorithm}
The IVFPQ algorithm\cite{jegou_product_2011} has two phases: index building and search.
The process of the index building phase is as follows.
\begin{enumerate}
    \item Split the dataset into some clusters, for instance by $k$-means
    \item In each cluster, calculate the residual vectors between each vector in the cluster and the centroid of the cluster.
    \item Quantize the residual vectors $\mathbf{r}$ with grouping some elements (product quantization).
    By this process, the $d$-dimensional vector $\mathbf{r}$ are compressed to $s$-dimensional vectors $\mathbf{y}$ where each element is quantized to $2^p$ patterns ($s < d$ and $p$ is called ``PQ bit").
    The example for $s=d/2$ is as follows.
    \[
        \mathbf{r} = \begin{bmatrix}r_1 \\ r_2 \\ \vdots \\ r_d \end{bmatrix} =
        \begin{bmatrix}\begin{bmatrix}r_1 \\ r_2\end{bmatrix} \\ \vdots \\ \begin{bmatrix} r_{d-1} \\ r_d \end{bmatrix} \end{bmatrix} 
        \begin{matrix}\underset{\text{Quantize}}{\to} \\ \vdots \\ \underset{\text{Quantize}}{\to}\end{matrix}
        \begin{bmatrix}y_1 \\ \vdots \\ y_s \end{bmatrix} = \mathbf{y}
    \]
    \item Save the cluster belonging information, compressed dataset $\{\mathbf{y}_1, \mathbf{y}_2, \cdots, \mathbf{y}_N\}$, and the relation of $[r_t \cdots r_u]^\top\underset{\text{Quantize}}{\to} y_v$ that is called PQ codebook.
\end{enumerate}

Then, the process of the search phase for a given $d$-dimensional query vector $\mathbf{q}$ is as follows.

\begin{enumerate}
    \item Pick up one or more clusters whose centroids are relatively close to the query vector.
    \item In each cluster, calculate the residual vector $\mathbf{r}^{q}$ between the query vector and the centroid.
    \item Compute the square of L2-norm between each sub-vector in the PQ codebook and that of $\mathbf{r}^q$ in the corresponding position.
    By this process, we can refer $|[r_t \cdots r_u]^\top - [r^q_t \cdots r^q_u]^\top|^2$ from a PQ index $y_v$.
    We denote this reference process as a function $F(y_v)$.
    \item For each compressed dataset vector $\mathbf{y}$ in the cluster, compute
    \begin{equation}
        \label{eq:accumulation}
        R=\sum_{j=1}^s F(y_j).
    \end{equation}
    Since the residual of the residual vectors $\mathbf{r}$ and $\mathbf{r}^q$ corresponds to the residual of the original vector $\mathbf{x}$ and $\mathbf{q}$, the $R$ is the square of the L2-norm of it.
    \item Find the top-$k$ nearest vectors, (partially) sorting the $R$ for each vector $y$ in the clusters.
\end{enumerate}
By constructing the reference table for $F(y_v)$ at first, we can compute $R$ just by loading the fragments from memory and accumulating them.
We call this reference table the ``norm2 fragment lookup table".

\begin{figure}
    \centering
    \includegraphics[width=\linewidth]{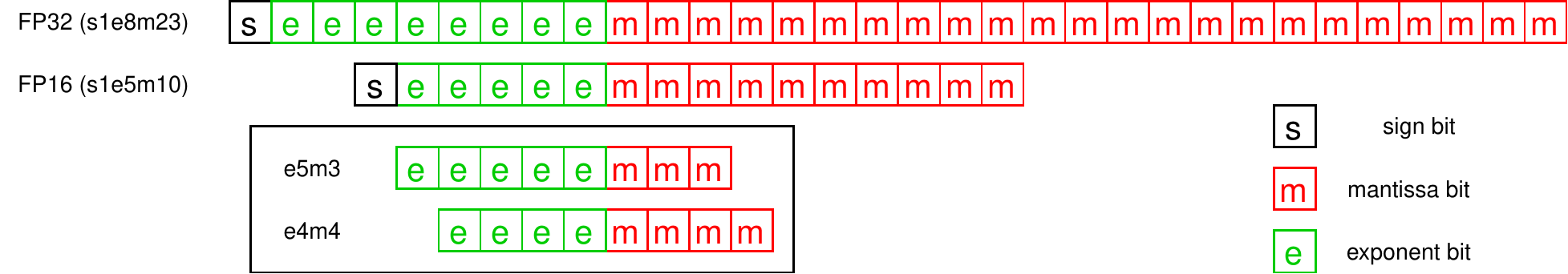}
    \caption{The comparison of our 8-bit floating point value formats and IEEE binary32/16.}
    \label{fig:fp}
\end{figure}
\begin{figure}
    \centering
    \includegraphics[width=\linewidth]{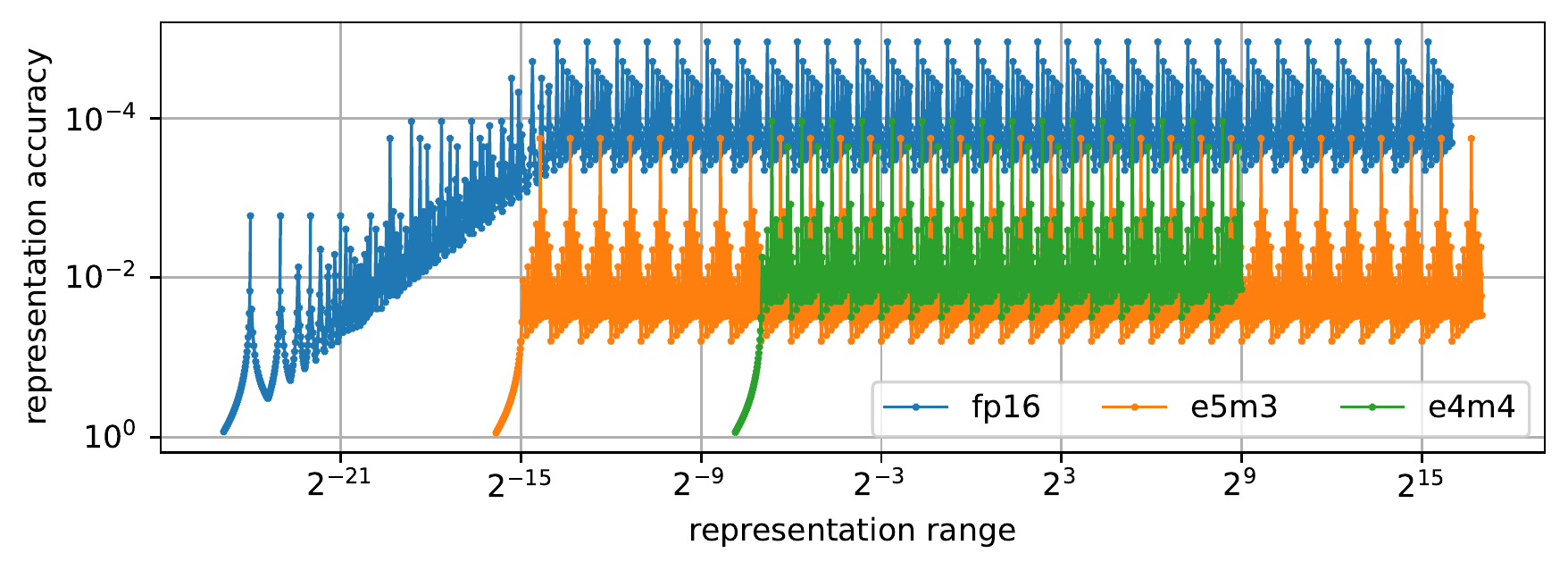}
    \caption{The representation range and accuracy of E5M3 and E4M4}
    \label{fig:fp8-representation}
\end{figure}
\subsection{Norm2 fragment lookup table on the shared memory}
In the implementation of the IVFPQ search phase on GPU, it is natural to make the norm fragment table on the shared memory and each thread computes the $R$ for different $\mathbf{y}$ in the same cluster.
When the PQ bit $p=8$, the number of patterns of $F(y_l)$ is $2^8=256$.
And when the return data type of the function is FP32 (4 bytes), each bank in the shared memory holds the elements in $256 / 32 [\text{=num banks}] / (4 [\text{=data type size}] / 4 [\text{=bank width}]) = 8$ lanes.
Therefore, $8-1 = 7$ bank conflicts can occur in the worst case.
We have profiled the number of bank conflicts in runtime and figured out that it degrades the throughput.
When using FP16 (2 bytes) for the return data type, only $4$ lanes are needed since the 1 bank can access 2 elements at once.
In this case, only $4-1 = 3$ bank conflicts can occur in the worst case.
So, how about using FP8 (1 byte)?

\section{Custom 8-bit floating point value format for IVFPQ search on GPUs}
\begin{figure}
\begin{minipage}[t]{0.48\hsize}
    \captionsetup{width=.95\linewidth}
    \includegraphics[width=\linewidth]{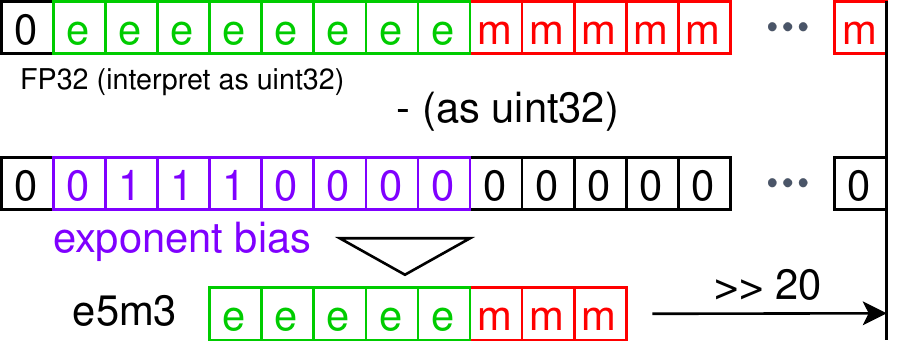}
    \caption{Conversion from FP32 to {\tt e5m3}.}
    \label{fig:from-fp32}
\end{minipage}
\hfill
\begin{minipage}[t]{0.48\hsize}
    \captionsetup{width=.95\linewidth}
    \includegraphics[width=\linewidth]{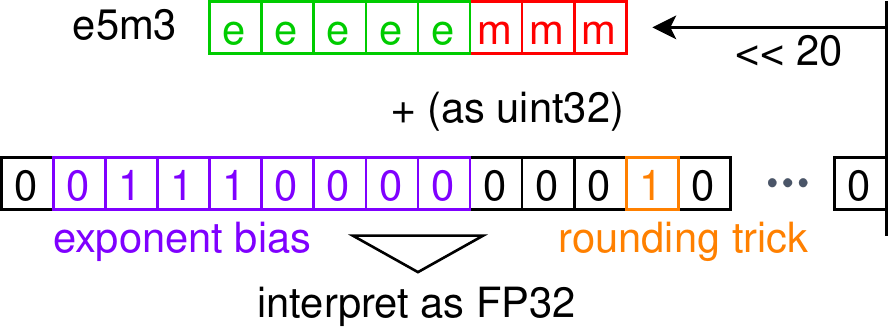}
    \caption{Conversion from {\tt e5m3} to FP32.}
    \label{fig:to-fp32}
\end{minipage}
\end{figure}

We designed two 8-bit floating point value formats for data storage to reduce the bank conflict: {\tt e5m3} and {\tt e4m4}, where the numbers of exponent bits are 5 and 4, and that of the mantissa are 3 and 4, respectively, as shown in Figure \ref{fig:fp}.
Since the return values of $F(y_v)$ are always zero or positive, we do not need a sign bit.
This feature allows us to increase the exponent or mantissa bits and easily convert them from/to FP32 format using two instructions: shifting the floating point value bitstring and adjusting the exponent bias.
We show the representation range and accuracy of the formats in Figure \ref{fig:fp8-representation}.
The {\tt e5m3} can represent the whole range of FP16 normalized numbers, but the accuracy is lower.
The features of the format and implementation are as follows:
\begin{itemize}
    \item Only for data storage. No computation between the formats.
    \item No sign bit since the values we want to store are always positive.
    \item Small overhead for converting from/to FP32.
    \item No support for non-normalized number and Inf/NaN.
\end{itemize}
We show the conversion operations from/to FP32 in Fig. \ref{fig:from-fp32} and Fig. \ref{fig:to-fp32}.

\section{Evaluation}
\begin{figure}
    \centering
    \includegraphics[width=\linewidth]{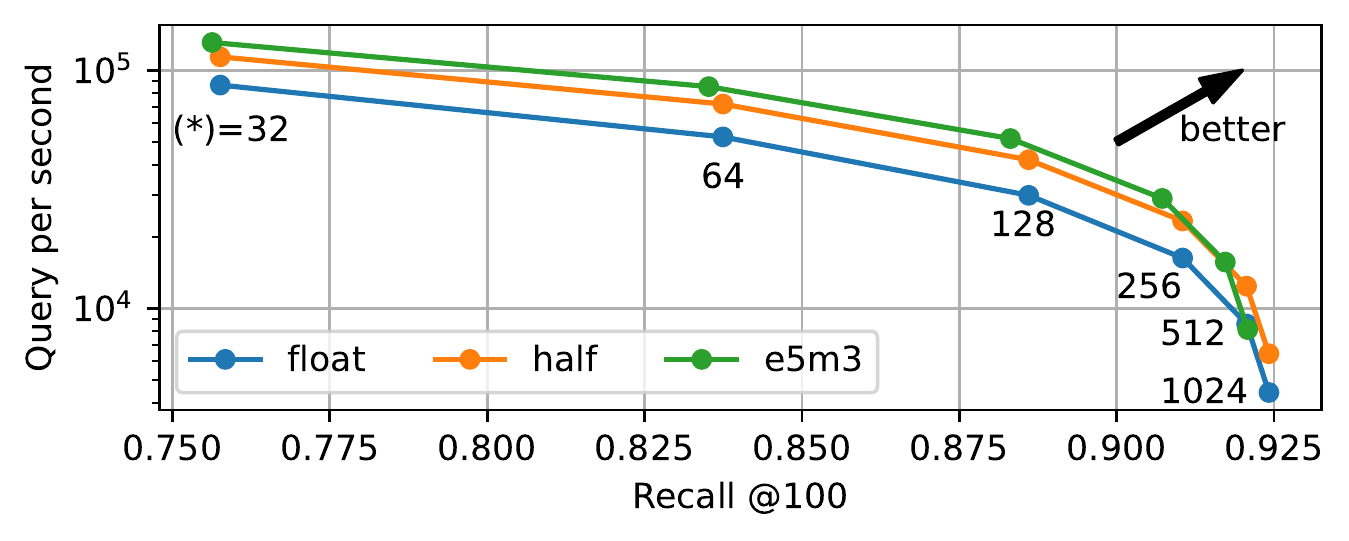}
    \caption{The recall and throughput for BIGANN-100M dataset when using {\tt float}, {\tt half} and {\tt e5m3} for the data type of norm2 fragment lookup table. The (*) is the number of clusters picked up in the first stage of the search phase.}
    \label{fig:bigann-eval}
\end{figure}

\begin{figure}
    \centering
    \includegraphics[width=\linewidth]{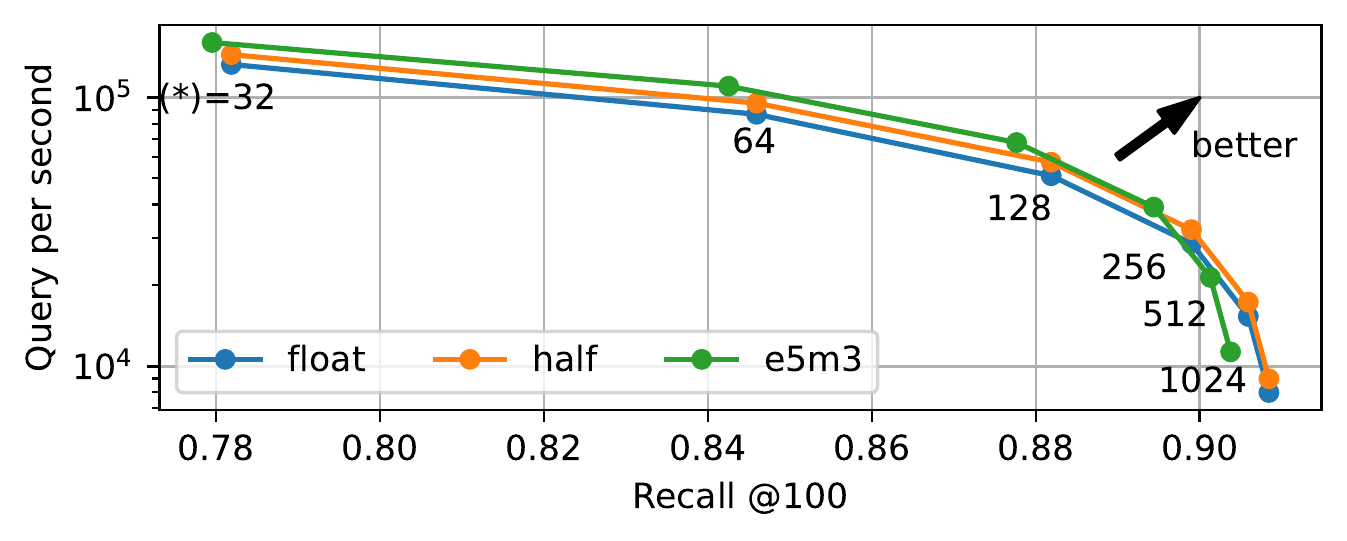}
    \caption{The recall and throughput for DEEP-100M when using {\tt float}, {\tt half} and {\tt e5m3} for the data type of norm2 fragment lookup table. The (*) is the number of clusters picked up in the first stage of the search phase.}
    \label{fig:deep-eval}
\end{figure}

We have evaluated the recall and throughput when using {\tt e5m3} for the data type of norm2 fragment lookup table on shared memory.
Our base implementation is LIBCUANN\cite{simhadri_results_2022}.
The datasets for the evaluation are BIGANN-100M and DEEP-100M, where $d=128$ and $96$, respectively.
We show the recall and throughput in Figure \ref{fig:bigann-eval} and \ref{fig:deep-eval}.
Using the {\tt e5m3}, we achieved higher throughput in both BIGANN-100M and DEEP-100M than FP32 and FP16 with a little recall degradation.
Although we also evaluate {\tt e4m4}, the recall and throughput are at the same level as {\tt e5m3}.

\section{Conclusion}
We have designed custom 8-bit floating point value formats for improving the IVFPQ throughput by reducing the shared memory bank conflict.
We have applied it to IVFPQ and improved the throughput with a bit of recall degradation.
\bibliographystyle{plain} 
\bibliography{references}
\end{document}